\documentstyle[buckow]{article}

\newcounter{multieqs}

\newcommand{\bq}{\begin{equation}}
\newcommand{\fq}{\end{equation}}
\newcommand{\bqr}{\begin{eqnarray}}
\newcommand{\fqr}{\end{eqnarray}}

\newcommand{\be}{\begin{equation}}
\newcommand{\ee}{\end{equation}}
\newcommand{\eq}[1]{(\ref{#1})}

\newcommand{\bm}[1]{\mbox{\boldmath $#1$}}

\def\bd{\begin{document}}
\def\ed{\end{document}}
\def\nn{\nonumber}
\def\bea{\begin{eqnarray}}
\def\eea{\end{eqnarray}}
\let\bm=\bibitem
\let\la=\label

\def\npb#1#2#3{Nucl. Phys. {\bf{B#1}} #3 (#2)}
\def\plb#1#2#3{Phys. Lett. {\bf{#1B}} #3 (#2)}
\def\prl#1#2#3{Phys. Rev. Lett. {\bf{#1}} #3 (#2)}
\def\prd#1#2#3{Phys. Rev. {D \bf{#1}} #3 (#2)}
\def\cmp#1#2#3{Comm. Math. Phys. {\bf{#1}} #3 (#2)}
\def\cqg#1#2#3{Class. Quantum Grav. {\bf{#1}} #3 (#2)}
\def\nppsa#1#2#3{Nucl. Phys. B (Proc. Suppl.) {\bf{#1A}}#3 (#2)}
\def\ap#1#2#3{Ann. of Phys. {\bf{#1}} #3 (#2)}
\def\ijmp#1#2#3{Int. J. Mod. Phys. {\bf{A#1}} #3 (#2)}
\def\rmp#1#2#3{Rev. Mod. Phys. {\bf{#1}} #3 (#2)}
\def\mpla#1#2#3{Mod. Phys. Lett. {\bf A#1} #3 (#2)}
\def\jhep#1#2#3{J. High Energy Phys. {\bf #1} #3 (#2)}
\def\atmp#1#2#3{Adv. Theor. Math. Phys. {\bf #1} #3 (#2)}

\def\a{\alpha}          \def\da{{\dot\alpha}}
\def\b{\beta}           \def\db{{\dot\beta}}
\def\c{\gamma}  \def\C{\Gamma}  \def\cdt{\dot\gamma}
\def\d{\delta}  \def\D{\Delta}  \def\ddt{\dot\delta}
\def\e{\epsilon}                \def\vare{\varepsilon}
\def\f{\phi}    \def\F{\Phi}    \def\vvf{\f}
\def\h{\eta}
\def\k{\kappa}
\def\l{\lambda} \def\L{\Lambda}
\def\m{\mu}     \def\n{\nu}
\def\p{\pi}     \def\P{\Pi}
\def\r{\rho}
\def\s{\sigma}  \def\S{\Sigma}
\def\t{\tau}
\def\th{\theta} \def\Th{\Theta} \def\vth{\vartheta}
\def\X{\Xeta}

\def\cA{{\cal A}} \def\cB{{\cal B}} \def\cC{{\cal C}}
\def\cD{{\cal D}} \def\cE{{\cal E}} \def\cF{{\cal F}}
\def\cG{{\cal G}} \def\cH{{\cal H}} \def\cI{{\cal I}}
\def\cJ{{\cal J}} \def\cK{{\cal K}} \def\cL{{\cal L}}
\def\cM{{\cal M}} \def\cN{{\cal N}} \def\cO{{\cal O}}
\def\cP{{\cal P}} \def\cQ{{\cal Q}} \def\cR{{\cal R}}
\def\cS{{\cal S}} \def\cT{{\cal T}} \def\cU{{\cal U}}
\def\cV{{\cal V}} \def\cW{{\cal W}} \def\cX{{\cal X}}
\def\cY{{\cal Y}} \def\cZ{{\cal Z}}

\def\ua{\underline{\alpha}}
\def\ub{\underline{\phantom{\alpha}}\!\!\!\beta}
\def\uc{\underline{\phantom{\alpha}}\!\!\!\gamma}
\def\um{\underline{\mu}}
\def\ud{\underline\delta}
\def\ue{\underline\epsilon}
\def\una{\underline a}\def\unA{\underline A}
\def\unb{\underline b}\def\unB{\underline B}
\def\unc{\underline c}\def\unC{\underline C}
\def\und{\underline d}\def\unD{\underline D}
\def\une{\underline e}\def\unE{\underline E}
\def\unf{\underline{\phantom{e}}\!\!\!\! f}\def\unF{\underline F}
\def\unm{\underline m}\def\unM{\underline M}
\def\unn{\underline n}\def\unN{\underline N}
\def\unp{\underline{\phantom{a}}\!\!\! p}\def\unP{\underline P}
\def\unq{\underline{\phantom{a}}\!\!\! q}
\def\unQ{\underline{\phantom{A}}\!\!\!\! Q}
\def\unH{\underline{H}}

\def\As {{A \hspace{-6.4pt} \slash}\;}
\def\Ds {{D \hspace{-6.4pt} \slash}\;}
\def\ds {{\del \hspace{-6.4pt} \slash}\;}
\def\ss {{\s \hspace{-6.4pt} \slash}\;}
\def\ks {{ k \hspace{-6.4pt} \slash}\;}
\def\ps {{p \hspace{-6.4pt} \slash}\;}
\def\pas {{{p_1} \hspace{-6.4pt} \slash}\;}
\def\pbs {{{p_2} \hspace{-6.4pt} \slash}\;}

\def\Xh{\hat{X}}
\def\ah{\hat{a}}
\def\xh{\hat{x}}
\def\yh{\hat{y}}
\def\ph{\hat{p}}

\def\pt{\tilde{p}}

\def\d{\delta}\def\D{\Delta}\def\ddt{\dot\delta}

\def\pa{\partial} \def\del{\partial}
\def\xx{\times}

\def\trp{^{\top}}
\def\inv{^{-1}}
\def\dag{^{\dagger}}
\def\pr{^{\prime}}

\def\rar{\rightarrow}
\def\lar{\leftarrow}
\def\lrar{\leftrightarrow}

\newcommand{\0}{\,\!}      
\def\one{1\!\!1\,\,}
\def\im{\imath}
\def\jm{\jmath}

\newcommand{\tr}{\mbox{tr}}
\newcommand{\slsh}[1]{/ \!\!\!\! #1}

\def\vac{|0\rangle}
\def\lvac{\langle 0|}

\def\hlf{\frac{1}{2}}
\def\ove#1{\frac{1}{#1}}

\def\Box{\square}
\def\bz{\bar{z}}
\def\ZZ{\mathbb{Z}}
\def\CC#1{({\bf #1})}
\def\bcomment#1{}
\def\bfhat#1{{\bf \hat{#1}}}
\def\VEV#1{\left\langle #1\right\rangle}

\def\ft#1#2{{\textstyle{{\scriptstyle #1}\over {\scriptstyle #2}}}}
\newcommand{\ex}[1]{{\rm e}^{#1}} \def\ii{{\rm i}}
\def\Dth{{\Delta_\th}}

\begin{document}

\hfill{DCPT-01/17 } {hep-th/0102164}
 

\def\titleline{
Multiloop Noncommutative Open String Theory 
\newtitleline
and their QFT limit.
}
\def\authors{
Chong-Sun Chu\1ad and Rodolfo Russo\2ad
}
\def\addresses{
\1ad Center for Particle Theory,
Department of Mathematical Sciences, \\ University of Durham, DH1 3LE, UK \\
\2ad Laboratoire de Physique Th\'eorique de
l'Ecole Normale Sup\'erieure,  \\
24 rue Lhomond, {}F-75231 Paris Cedex 05, France \\
\vskip .1cm
\sffamily{chong-sun.chu@durham.ac.uk, rodolfo.russo@lpt.ens.fr}
}
\def\abstracttext{
The multiloop amplitudes for open bosonic string in presence of a 
constant $B$-field are derived from first principles. The basic
ingredients of the construction are the commutation relations for the
string modes and the Reggeon vertex describing the interaction among
three generic string states.  The modifications due to the presence of
the $B$--field affect non--trivially only the zero modes. This makes it
possible to write in a simple and elegant way the general expression
for multiloop string amplitudes in presence of a constant $B$-field.
The field theory limit of these string amplitudes is also considered.
We show that it reproduces exactly the Feynman diagrams of
noncommutative field theories. Issues of UV/IR are briefly discussed.
}
\large
\makefront

\section{Multiloop NCOS and Reggeon Formalism}

The first basic ingredient for the construction of string
amplitudes in the operator formalism are the commutation 
relations for the string modes. As usual, these commutation 
relations can be derived from the the tree level world--sheet 
action. We consider an open string
ending on a D-brane in presence of a constant $B$-field.
The open string mode expansion is
\be \label{mode1}
X^\m(\tau,\sigma) =x_0^\m + 2\a' (p_0^\m \tau -  p_0^\n 
F_\n{}^\m \sigma) +
\sqrt{2\a'} \sum_{n\neq 0} {e^{-\ii n\tau} \over n}
\bigl(\ii a^k_n \cos n\sigma -   a_n^\n F_\n{}^\m \sin 
n\sigma \bigr)~,
\ee
where $F =B-dA$ is the modified Born-Infeld field strength.
Canonical quantization yields the following commutation
relations~\cite{CH1} 
\bea
&[a_n^\m, x_0^{\n}]=[a_n^\m, p_0^\n] = [p_0^\m, p_0^\n]=0,
\label{cr1} \\
&[a_m^\m, a_n^\n]= mM^{-1\m\n}\d_{m+n},
\quad [x_0^\m, p_0^\n]=i M^{-1\m\n}, \quad 
[x_0^\m,x_0^\n]= i \,\Theta^{\m\n}, \label{cr2}
\eea
where $M_{\m\n} = g_{\m\n} - (F g^{-1} F)_{\m\n}$ is the 
open string metric
and 
\be \label{openmetric}
(M^{-1})^{\m\n} = \Big( \frac{1}{g+ F} g \frac{1}{g- 
F}\Big)^{\m\n}, \quad
\Theta^{\m\n} = 2\pi\a' \Big( \frac{1}{g+ F} F \frac{1}{g- 
F}\Big)^{\m\n}
\ee
are the symmetric and antisymmetric part of the matrix
$ (\frac{1}{g+ F})^{\m\n}= (M^{-1})^{\m\n}- 
\frac{\Theta^{\m\n}}{2\pi
\a'}$. Although these commutation relations were derived at 
the tree level, they are valid at all loops \cite{crs,dolan} and 
therefore one can use them directly to construct the higher loop 
string amplitudes. Due to the limitation of vertex operator formalism
to go beyond 1-loop, it is necessary to introduce the Reggeon vertex  
formalism for string amplitudes. 

The basic object in the Reggeon formalism is the 3-Reggeon vertex which
describes, at the tree level, the interaction among three generic string
states~\cite{dss}
\be \label{V30} V_{3;0}^\th(\zeta) = \int \!\! 
\frac{dp}{\sqrt{\det M}} \;\langle p, 0;
q=3| : {\rm e}^{\left\{\oint_0 dz (-X^v(\zeta+z)\partial_z X(z)-
c^v(\zeta+z)b(z) +b^v(\zeta+z) c(z) )\right\}}:~. 
\ee
Here the bra indicates the vacuum of the emitted string with 
momentum $p$ and the label $q=3$ specifies the ghost number.
$X^v$ is the virtual propagating string. 
$X$ and $X^v$ both have an
expansion of the form \eq{mode1} with commutations 
relations 
given by \eq{cr1}, \eq{cr2}, while the virtual and the
external strings simply commute among themselves.
Notice that that \eq{V30} is almost identical to the standard
3-Reggeon vertex in the trivial background $B=0$. The only
modification with the respect to the usual case is the appearence of a
factor of $\det M$ in the measure of momentum integrals which is due
to the fact that the open string metric is flat, but non
trivial\footnote{We thank O. Andreev for pointing out to us this fact.
  In Ref.~\cite{crs} this factor was absent because the noncommutative
  Reggeon-Vertices and amplitudes were written in terms of rescaled string
  coordinates, as we do in the following Section 3.1.}. More generally
one has to modify the usual ($B=0$) normalizations every time the
volume of the space seen by the open strings appears: for instance,
the open string vacuum has to be normalized as $\langle 0| 0 \rangle
=\sqrt{\det M}~V$ and, thus, the generic momentum state satisfies
$\langle p| p' \rangle =\sqrt{\det M}~ \delta^{d} (p-p')$.  However,
even if \eq{V30} is formally unchanged, it contains a non-trivial
dependence on $B$ through the mode expansion \eq{mode1} and the new
commutation relations \eq{cr1}, \eq{cr2}.
It should also be stressed that the zero-mode $x_0$ (or $y_0$ if the
interaction is at $\s=\pi$) appears only in the expansion of the virtual
string and this is the only source of the non-trivial dependence on
$\Theta$. We will employ the usual physical states having ghost number
$1$; thus, the $(b,c)$ system is not affected by the background field $F$
and one recovers the well-known results for the ghost contributions.
Because of this, in what follows, ghosts will no longer be mentioned, and
we will focus only on the $F$-dependent modifications coming from the
orbital part. 

The tree level $N$-Reggeon vertex is obtained by simply multiplying $N$
$3$-Reggeon vertices in different positions $\zeta$, but with the common
propagating string $X^v$.  Finally one takes the vacuum expectation value
in the Hilbert space of the propagating string in order to obtain a
symmetric object in the $N$ external states. The new $\Th$-dependent part
comes when one collects together the zero mode factors. In particular, if
the external legs of all the original 3-Reggeon vertices are emitted from
the border $\s=0$, one obtains the new phase factor 
\be 
  e^{i p^1 x_0} \cdots e^{i p^N x_0} = e^{-\frac{i}{2} \sum_{r <s}^N
  {p}^{r}\Th {p}^{s}}, 
\ee 
where momentum conservation has been used. As a result, we obtain the
$N$-Reggeon vertex with all legs emitted from the $\s=0$ border 
\be
\label{tree} V_{N;0}^\Th = \sqrt{\det M}~ V_{N;0}^0 \exp (-\frac{i}{2} 
\sum_{i<j}^N
{p}^{i} \Th {p}^{j}), 
\ee 
where ${p}^{i}_\m$ is the momentum {\it operator} of the $i$-th leg, in
the direction flowing towards the boundary. Here $V_{N;0}^0$ indicates
the $N$-Reggeon vertex derived for the usual ``commutative'' case
$F=0$ in \cite{sc2}. This vertex is a bra in the direct product of the
$N$ distinct Fock spaces for the external strings; like the
$3$-Reggeon vertex, it gives the scattering amplitude when the
external legs are saturated with physical states. Note that the
momentum dependent phase factor is exactly the same modification as
the one introduced by the Moyal $*$-product in the tree level vertex
of a noncommutative field theory \cite{filk}.

The $h$-loop $N$-Reggeon vertex in the presence of a constant $F$-field
can be constructed by sewing together pairs of legs in a tree level
($N+2h$)-Reggeon vertex. The sewing is achieved by using the BRST
invariant operator $P(x)$, which is a function of $L_0$ 
and $L_{\pm 1}$ and of the ghosts. Since $L_n$ does not involve the zero
modes $x_0$, we conclude that the string propagator $P(x)$ and the sewing
procedure are not modified by the presence of $F$-field. Therefore the
only new feature for $\Th  \neq 0$ is in the zero modes part of the
Reggeon vertex. And we obtain~\cite{crs}
\be\label{Vloop}
V_{N;h}^\Th = \sqrt{\det M}~ \widetilde{V}_{N;h}^{0}
\prod_{I=1}^{h} \int \frac{dp^I}{\sqrt{\det M}} 
\exp\left(
\frac{1}{2} \sum_{I,J =1}^h p^I_\m A_{IJ}^{\m\n} p^J_\n +
\sum_{I =1}^h B_I^\m p^I_\m + C
\right),
\ee
where $\widetilde{V}_{N;h}^{0}$ \cite{phi,sc2}
contains the ghost contribution and
only the nonzero mode piece of the orbital part of 
$N$-Reggeon vertex
in absence of background. Here
\be
A_{IJ}^{\m\n} ={A^0}_{IJ}^{\m\n} - i \Th^{\m\n}\cJ_{IJ},  
\quad
B_I^\m = {B^0}_I^\m - i \Th^{\m \n} {P}_{I\n},\quad 
C= C^0 - \frac{i}{2} \sum_{i<j}^N {p}^i \Th {p}^j,  
\label{abc}
\ee
where $p_I$ are the loop momenta with loop indices $I=1, \cdots h$;  
$p_i$ are the external momenta with legs labelled by $i= 1,\cdots, N$;
$\cJ_{IJ}$ is the intersection matrix for the internal loops; $P_I$ is the
sum of the external momenta leaving the $I^{th}$ loop, and
$\mu,\nu$ stands for the spacetime indices.
$A^0, B^0, C^0$ are independent of $\Th$ and are given by 
\cite{phi},
\bea
{A^0}_{IJ}^{\m\n} &=&  2\a'(2\pi i \t_{IJ}) (M^{-1})^{\m\n} 
,\quad\quad\quad
{B^0}_I^\m =
\frac{1}{2\pi}  \sum_{i=1}^N \oint_{0} dz \del X^{(i) 
\m}(z)
\int_{z_0}^{V_i(z)} \omega_I ~ ,     \label{abc0}  \\
C^0 &=& -\frac{1}{2}\sum_{i=1}^N
\oint_{0} dz \del X^{(i)}(z) p_0^{(i)} \ln V'_i(z)  \nn \\
&&+ \frac{1}{2\a'} \sum_{i<j}^N \oint_{0} dz \oint_{0} dy 
\del X^{(i)}(z)
\ln [V_i(z)-V_j(y)] \del  X^{(j)}(y) \nn\\
&&+ \frac{1}{4\a'} \sum_{i,j=1}^N \oint_{0} dz \oint_{0} dy
\del X^{(i)}(z) \ln \left(\frac{E(V_i(z),V_j(y)) 
}{V_i(z)-V_j(y) }\right)
\del  X^{(j)}(y),  \nn
\eea
where, due to \eq{cr1} and \eq{cr2}, all indices are contracted by the
open string metric $M^{-1}$.  Here $V_i(z)$ is chosen to satisfy
$V_i^{-1}(z)=0$ for $z=z_i$. $\omega_I$ is the normalized Abelian
differential and $\t_{IJ}$ is the period matrix and $E(z,w)$ is the prime
form. Their explicit expressions in term of the Schottky parameters can be
found in \cite{phi}.

Note that all the dependence in $\Th$ is localized in the zero modes loop
momentum integration. Carrying out the loop momentum integration, one
obtains finally 
\be \label{stringmaster} V_{N;h}^\Th = \left[\sqrt{\det M}\right]^{1-h}
\widetilde{V}_{N;h}^{0} \frac{1}{\sqrt{\det \frac{-A}{2} }} \exp
(-\frac{1}{2} B^T A^{-1} B +C) , 
\ee 
where the determinant is taken over
the space of Lorentz and loop indices $(\mu I)$. The effects of $\Th$ are
summarized elegantly in \eq{abc}.

\vspace{10pt}

\underline{Modifications of $\Th$ to the string amplitude}

As an illustration of how $\Th \neq 0$ modify the string amplitude,  
it is instructive to recall the explicit form of the higher loop 
string amplitude  for $\Th=0$. For example, the string
amplitude  for $M$-gauge bosons can be obtained by 
saturating the  $V_{M;h}^{\Th=0}$
with $M$ external gluon states. Using  the explicit formula \eq{abc0},
one obtains 
\bea\label{gluonmaster0}
&& A^{(h)}_M (p_1,\ldots,p_M) =
C_h {\cal N}_0^M \int [dm]_h 
\frac{\prod_{i=1}^M d\rho_i/\rho_i}{dV_{abc}} \prod_{i<j} 
~\exp \left(\sum_{i<j} 2\alpha' p_i^\m p_j^\n 
(G_{\m\n}(\r_i,\r_j)
\right)
\nn \\
&  \times & \left[ \exp  \sum_{i\not=j} \left(
\sqrt{2\a'}\, 
\e_i^\mu \,\rho_i \del_{\r_i}  G_{\m\n}(\r_i,\r_j) p_j^\n 
+\, {1\over 2} \e_i^\mu \, \rho_i \rho_j 
\del_{\r_i}\partial_{\r_j} G_{\m\n} (\r_i,\r_j) \e_j^\n 
\right) \right],  
\eea
where only terms linear in each polarization
should be kept.
In the general case of having other external states,
one need to change correspondingly
the form of the second line in \eq{gluonmaster0},
however the first line is universal.
Here in \eq{gluonmaster0} 
$ dV_{abc} =d \psi_a d \psi_b d \psi_c 
(\psi_a-\psi_b)^{-1}(\psi_a-\psi_c)^{-1}(\psi_b-\psi_c)^{-1}$ 
is the projective invariant volume element, 
where $\psi_a,\psi_b,\psi_c$ are any three of the $M$ punctures
$\rho_i$ or of the $2h$ fixed points $\xi_{\mu}$, $\eta_{\mu}$ 
of the generators of the Schottky group.  
The normalization constants 
$\cN_0$ and $C_h$ are given by \cite{russo1} and one can check that
their functional form is unmodified by the presence of $F$
\be \label{CN}
\cN_0 = \sqrt{2} g_{o} (2\a')^{\frac{d-2}{4}}, \quad\quad
C_h = \frac{1}{(2\pi)^{dh}} g_{o}^{2h-2} \frac{1}{(2\a')^{d/2}} ,
\ee
where $g_{o}$ is the open string coupling.
To relate the open string coupling $g_o$ with the 
closed string coupling $g_s$, one can  
study the factorization of the annulus diagrams. In
Ref.~\cite{frau1997} the relation between the two couplings is
explicitely written in terms of the volume $V$ seen by the open
strings (see Eq.~(3.10)). As we have already said, in the presence of
a background $F$-field $V$ has to be replaced by $\sqrt{\det  M}~V$. 
Thus from the result of~\cite{frau1997} one can derive the
$F$-dependence of $g_o$  
\be
g_o \propto (\det M)^{1/8} g_s\sim \Big[\det(1+F)\Big]^{1/4} g_s~,
\ee 
which gives an implicit dependence on $F$ to all of the above open string
quantities.
This relation was derived in \cite{sw1} by looking at the Born-Infeld actions.
Here we give a string derivation of this relation within the operator
formalism. 
The measure factor $[dm]_h$ for $\Th=0$  is given by
\be \label{measureh}
[dm]_h = 
\prod_{\mu=1}^{h} \left[ \frac{dk_\mu d \xi_\mu d \eta_\mu}{k_\mu^2
(\xi_\mu - \eta_\mu)^2} ( 1- k_\mu )^2 \right]   \label{lm:meas} 
\left[\det \left( - 2\pi i\a' \tau_{IJ} \right) \right]^{-d/2} 
\prod_{\alpha}{}' \left[ \prod_{n=1}^{\infty} ( 1 - k_{\alpha}^{n})^{-d}
\prod_{n=2}^{\infty} ( 1 - k_{\alpha}^{n})^{2} \right]    .
\nn
\ee 
where $k_{\mu}$ are the multipliers and $\xi_{\mu}$, $\eta_{\mu}$ are
the fixed points of the  Schottky  generators
of the surface ~\cite{phi}; 
the primed product over $\alpha$ denotes a product over conjugacy classes 
of elements of the Schottky group, where only elements that cannot be 
written as powers of other elements must be included. Three of $2 h + M$
parameters $\xi_\mu$, $\eta_\mu$ and $\rho_i$ can be fixed using an overall 
projective invariance of the amplitude. The fixing of this invariance
introduces the projective invariant volume element $dV_{abc}$ in 
\eq{gluonmaster0}. Including the ``multipliers'' $k_\mu$, one is left 
with $3 h - 3 + M$ variables, the correct number of independent moduli 
for a Riemann surface of genus $h$ with $M$ punctures. 

When $\Th$ is turned on, the shift in \eq{abc} for $A, B, C$ modify
the exponent of \eq{stringmaster}. This part of the modification 
may be summarized in terms of a modified Green function depending on $\Th$. 
In addition, the shift in $A$ gives rise to  
a $\Th$ dependent measure factor $ \sqrt{\det (-A/2)} $ 
in \eq{stringmaster}. 
Notice that in the particular case of the $h$-loop bubble backbone
with no intersections $\cJ=0$, the $\det M$ factors cancel out
and the $F$-dependence is completely determined by that of
$g_0$ and by the universal factor $\sqrt{\det M}$ present in~(\ref{Vloop}),
\be
V_{0;h}^\Th = [{\det( 1+ F)}]^{\frac{1+h}{2}} \;\times~V_{0;h}^{0}. 
\ee
For $h=1$, this agrees with the computation of \cite{callan}, and, in
the general case, we recover the result of~\cite{Andreev}.

\section{One-loop Amplitudes and Green Function}

Having the general form \eq{stringmaster} 
of the multiloop Reggeon vertex, it is
straightforward  to write down explicitly the 1-loop 
amplitude and
extract the 1-loop Green function.

First, there is  no modification to
$A^0$ and the modification to $C^0$ is
simply the usual field theory
Filk phase. The 1-loop measure is given by
\be \label{dm1}
[dm]_1\frac{\prod_{i=1}^M d\rho_i/\rho_i}{dV_{abc}}  
= \prod_{i=2}^M \frac{d\rho_i}{\rho_i}
\frac{dk}{k^2} 
\left( - \frac{\ln k}{2} \right)^{-d/2}\prod_{n=1}^\infty(1- k^n)^{2-d} 
\ee
where $k$ is the multiplier of 
the annulus in the Schottky representation. We have exploited the
projective invariance to  fix 
$ \rho_1 = 1, \quad \eta=0, \quad \xi = \infty $
in \eq{dm1}. 
Since the measure is not modified, the 1-loop
amplitude can be written in the familiar form \eq{gluonmaster0} 
with all the $\Th$-dependence
taken care of by the Green function.
Thus we concentrate on the effects from the 
shift
of ${B^0}$. The one-loop  Abelian differential is
$ \omega = dz /z$. With the convenient choice of $V_i(\r) = 
\r+ \r_i$,
we have
\be
 {B^0}_I^\m =
\sum_{r\in I_1 \cup I_2} \cB^{r\,\m} (\r) \ln|\r+\r_r| 
\quad \mbox{with} \quad 
\cB^{r\,\m}(\r) = \frac{1}{2\pi i} \oint_0 d\r \del 
X^{(r)\m}(\r),
\ee
where $\cB^{r\m}$
is an operator in which the integration is carried out on 
any function
that multiply it on the right. One obtains finally
\bea \label{higher}
\frac{-1}{2A} B^2 &=& 2 \a' \cdot \frac{1}{8{\a'}^2 \ln k}
\Big( \sum_{\stackrel{r<s}{r,s\in I_1\cup I_2}}
\cB^r(\r) \cB^s(\r') \ln^2 \left|\frac{\r+\r_r}{\r'+\r_s}\right| \\
&+& \frac{i}{\a'} \sum_{\stackrel{r\in I_1}{s\in I_2}}
\cB^r(\r) \Th \cB^s(\r') \ln |(\r+\r_r)(\r'+\r_s)| +
\frac{1}{4 \a'^2} \sum_{\stackrel{r\in I_1}{s\in I_2}}
\cB^r(\r) \Th^2 \cB^s(\r')
\Big), \nn
\eea
where $I_1, I_2$ denote the two boundaries of the annulus. The second
line contains all the $\Th$-dependence of the one-loop $N$-Reggeon
vertex $V_{N;1}^\Th$.

Written in terms of the Schottky
representation of the annulus, the planar and nonplanar open string
Green functions are 
\bea \label{p1}
&& G_P^{\m\n}(\r,\r') = I_0^P (M^{-1})^{\m\n} -  
\frac{\ii\Th^{\m\n}}{4 \a'}
\epsilon(\r-\r'), \\
&& G_{NP}^{\m\n}(\r,\r') = I_0^{NP} (M^{-1})^{\m\n}
+ \frac{(\Th^{2})^{\m\n}}{ 8 \a'^2}\frac{1}{\ln k}
\pm  \frac{\ii\Th^{\m\n}}{2 \a'} \frac{\ln|\r \r'|}{\ln k}, 
\label{np1}
\eea
where the $\Th$-independent piece is given by
\be \label{p2}
I_0^P(\r,\r') = \frac{ \ln^2 \r / \r'}{2 \ln k} +
\ln \left|\sqrt{\frac{\r}{\r'}} - 
\sqrt{\frac{\r'}{\r}}\right|
+  \ln \prod_{n=1}^{\infty}
\left|\frac{(1-k^n \r/\r')(1-k^n \r'/\r) }{(1-k^n)^2} 
\right|  
\ee
for the planar case, while for non--planar contractions one has
\be
I_0^{NP}(\r,\r') = \frac{ \ln^2 |\r / \r'|}{2 \ln k} +
\ln (\sqrt{\frac{|\r|}{|\r'|}} + \sqrt{\frac{|\r'|}{|\r|}} 
) + \ln
\prod_{n=1}^{\infty} \left|\frac{(1+k^n |\r/\r'|)(1+k^n 
|\r'/\r|)
}{(1-k^n)^2} \right|~.
\ee
In the last term of the nonplanar Green function in \eq{np1}, positive
sign is taken when $\r >0, \r' <0$ and negative sign is taken for the
opposite case $\r' >0, \r <0$.

\subsection{Remarks on the  Green function}
In the above, the open string amplitudes and the 
open string Green function 
in the presence of $B$-field were obtained directly using the basic 
commutation relations \eq{cr1}, \eq{cr2} and the Reggeon formalism.  
A different approach of using the  {\it closed} string Green function 
as input was  adopted at the one loop level 
\cite{nn1,nn2,BCR,shenker,hong} to calculate the open 
string amplitude. The idea is to obtain the  {\it open} string
Green function by letting the arguments of the closed string Green
function to approach the boundary and then use it as input 
(in, {\it e.g.} \eq{gluonmaster0}) to calculate the open string amplitude. 
However as discussed in \cite{BCR,crs}, 
there are ambiguities  associated with this approach. 

{\bf 1.} Approaching the boundary

We  first recall the procedure of \cite{sw1} in obtaining the boundary
correlator, in particular its antisymmetric part, from the bulk.
The boundary correlator obtained there is
\be \label{treeG}
\cG^{\m\n}(\t,\t'):= \langle X^{\m}(\t)  X^{\n}(\t')  \rangle 
= -\a' (M^{-1})^{\m\n} \ln (\t-\t')^2 
+ \frac{i}{2} \Th^{\m\n} \e(\t-\t'),
\ee
where $M^{-1}$ and $\Theta$ are given in \eq{openmetric}.
The antisymmetric piece was obtained from the term 
\be
\Th^{\m\n}\ln \frac{z- \bar{z}'}{\bar{z} -z'}
\ee
in the bulk correlator by letting $z, z'$ to {\it approach} the
boundary in an appropriate manner. Notice that restricting the
arguments $z ,z'$ directly to the real axis yields $\ln 1 = 2 n \pi i $
instead of the moduli dependent piece in \eq{treeG}. Note also that
there is no uniform shift one can perform to the bulk correlator so
that  \eq{treeG} is obtained  when restricted to the boundary. A
limiting procedure must be adopted. 
We also remark that  
radial ordering (on $|z|$) is employed for operators in the
bulk, while a time ordering on $\t$ is employed on the boundary. The
two orderings agree on the positive $\t$ axis, but is opposite to each
other on the negative axis.

Thus there is a certain ambiguity in the
antisymmetric part of \eq{treeG} relating
to the choice of the branch cut of the $ln$. This ambiguity can be
fixed easily at the tree level and the form  \eq{treeG} is
the correct one. 

{\bf 2.} Choices of closed string Green functions

As noted in \cite{BCR}, there is an
additional source of ambiguity at one loop:
there is a freedom in the  definition of the closed string  
Green function.

At the one loop level, it 
was first noted by \cite{callan} that it is not possible to
strictly impose on the Green function the same boundary condition 
as imposed on the string coordinates
($\s =0, \pi$)
\be
\del_\s X^\m  + \del_\tau X^\n F_\n{}^\m =0, \quad \m,\n= 0,1,\cdots, p,
\label{BC1}\\
\ee
and hence
there is a certain degree of freedom in the choice of what constraint
is satisfied by $G$. If one insists on 
\be \label{g1a}
\Delta G_{\m\n}(z,z') = -2 \pi \a' \eta_{\m\n} \delta(z,z')
\ee
then since $\oint \partial_\perp G ds = -2 \pi \a'$ that is fixed by the 
Gauss theorem, one obtains the following set of compatible boundary
conditions 
\be \label{g1b}
 \frac{\del}{\del r} G_{\m\n}(z,z') - 
\frac{i}{r}  \frac{\del}{\del \th} F_\m{}^\l G_{\l\n}(z,z')
= \left\{
\begin{array}{ll}
-\frac{\a'}{a} \b & \mbox{ at $r=a$},\cr
-\frac{\a'}{a} (1-\b) & \mbox{ at $r=b$},
\end{array}
\right.
\ee
where $\b$ is arbitrary. 
The choice $\b=0$ was first derived by
\cite{callan} using the method of image and was  used in \cite{hong}
in the computation of the tachyon amplitudes, while the choice $\b=1/2$ was
adopted in \cite{nn1}.  
One can also introduce a background
charge so that a Green function  
\be\label{g2a}
\Delta G_{\m\n}(z,z') = -2 \pi \a' \eta_{\m\n} \delta(z,z') +
\frac{2 \pi\a'}{A},
\ee
where $A = \pi (b^2 -a^2)$ is the area of the annulus, 
satisfying the boundary condition
\be \label{g2b}
 \frac{\del}{\del r} G_{\m\n}(z,z') - 
\frac{i}{r}  \frac{\del}{\del \th} F_\m{}^\l G_{\l\n}(z,z')
= 0 ,\quad r=a,b
\ee
can be constructed. This closed string Green function was used 
in \cite{nn2,BCR,shenker} to derive the open string Green function
and in computing the 2 point function for noncommutative photons.
The origin of this freedom in defining the Green
function was analyzed from the point
of view of the boundary state approach in \cite{BCR}, where
the closed string Green functions was derived from the closed string
worldsheet with boundary states inserted so as to create an open string
worldsheet. There it was noted that the freedom in the definition of
the closed string Green function is related to a freedom in
interpreting the tachyon amplitudes in terms of contributions from  
the closed string 
Green function and contributions from the self-contraction $\cC$.
A shifted 
\be
G'_{\m\n}(z,z')   = G_{\m\n}(z,z') + \cM_{\m\n}(z,z')
\ee
with a $\cM$ satisfying certain conditions \cite{BCR} 
gives the same closed string 
tachyon amplitude. The reason that $G$ and $G'$ can gives the same
tachyon amplitude is because of momentum conservation. For higher 
closed string states, the Green function may
also be contracted with other available quantum numbers like
polarization which doesn't have a conservation law. 
So the shift is indeed possible only for tachyon amplitudes. 
For the amplitude of higher massive closed string states, 
it is \eq{p1} and \eq{np1}  which are to be used. 
As was shown in \cite{BCR},
they can be obtained from an
appropriate limiting procedure 
of the bulk Green function that solves 
\eq{g2a} and \eq{g2b}. 
Other form of Green function gives incorrect amplitudes. 

This kind of ambiguity persists at higher loops. 
While it can be fixed easily at the tree level, it is
more subtle at the loop level. 
When one go to higher loop, there are many more boundaries and
the above mentioned ambiguities with the Green function will be much
harder  to resolve \cite{kiem}.

At tree and the 1-loop level, all the effects of $\Th$ can
be summarized in terms of a modified Green function. One may therefore
also use other approaches to obtain the Green function
and use it as input in, for example \eq{gluonmaster0}, to calculate the
string amplitude. However this is no longer the case for two and
higher loops.  As we mentioned above, 
there are  new modifications to the measure 
that is not presented at the tree and 1-loop level. We stress that this 
$\Th$-dependent modification to the measure 
cannot be obtained from the Green function approach. 
The advantage of the Reggeon operator formalism \cite{crs} is that one
obtains the string amplitude  \eq{stringmaster} 
in one step and there is no need to isolate a Green function from it,
which is where the ambiguities lie.

\section{Field Theory Limit}

The relations between string and field theory amplitudes have been
thoroughly studied since the early days of dual
models~\cite{scherk71}.  It turned out that string amplitudes contain
very precise information on various perturbative quantities of
different field theories. In fact, even using the simple bosonic
string as a starting point, it is possible to recover, with a suitable
definition of the low energy limit, the results of the usual Feynman
diagrams for scalar~\cite{fmr,mp}, Yang--Mills~\cite{fmr2} or
gravity~\cite{bddpr} field theories (see also the references in these
papers).
 
In this section we briefly review the basic steps allowing to derive
field theories amplitudes from string expressions like the one in
\eq{gluonmaster0}. In fact, the same procedure normally used to recover the
Feynman diagrams of commutative theories can be also applied to the
results derived in the previous sections. Of course the background
$F$-field of the string calculations is now related to the appearance
of the non commutative parameters $\Theta$ in the field
results. However, the algorithm one uses to perform the low energy
limit is essentially not affected by the presence of the 
$F$-field~\cite{nn1,nn2,BCR,shenker,hong}:
all important modifications related to the new feature of
noncommutativity are already encoded in the string result and in
particular in Eq.~\eq{abc}.

It is usually said that field theory results are recovered from string
amplitudes simply by taking the limit $\alpha' \to 0$. Of course, this
prescription has to be suitably interpreted to yield sensible answers.
First, $\alpha'$ is a dimensionful parameter and what the above limit
really means is that the typical energy scale of the external states
is very small in comparison with the string scale ($\a' p_i\cdot p_j
\to 0$). Moreover, it is not possible to blindly perform this limit on
the amplitudes.  In fact, string results are written as integrals over
the moduli space of the Riemann surface representing the world--sheet.
Also, the integrand contains various divergences which make the
limiting procedure delicate. However, the general idea underlying the
derivation of field amplitudes from string ones is to obtain also the
answer for field diagrams in an integral form. In fact, the
world--sheet moduli are strictly related to the usual Schwinger
parameters introduced in diagrammatic perturbative
computations. Moreover, only the corners of moduli space where the
integrand diverges contribute to the field theory limit.  Thus, a
single string amplitude decomposes in a sum of different contributions
coming from different corners of the full integration region. It turns
out that each of these terms encodes the result of all Feynman
diagrams of a given topology (for instance, in the Yang--Mills case
the string approach automatically sums the contribution of ghost and
gluon propagation).

Summarizing the low energy limit on string amplitudes is basically
done in three steps:
\begin{itemize}
\item First it is necessary to express in terms of string quantities
all the parameters appearing in the lagrangian of the field theory one
wants to reproduce. The dictionary between the two sets of parameters
can be easily derived by looking at the simplest diagram in both
theories; for instance, by matching the $3$-point amplitudes, one can
usually find the relation between the string and the field theory
coupling constants.
\item Then one has to focus on the different relevant corners of the
region of integration. In each case, the string moduli are transformed
into Schwinger parameters with a relation of the type $t = -\a' \ln
f(z)$, for some function $f$ which may depend on the details of the
worldsheet parameterization.
\item Finally one can send $\a'\to 0$, but has to keep all field
theory parameters finite (even if they are dimensionful, like the
Schwinger proper parameters)
\end{itemize}

\subsection{Taking the field theory limits} 
As it turns out, there are two ways one can derive the field theory
limit.  In \cite{sw1}, the Seiberg-Witten limit 
\be 
\a' \sim \e, \quad
g \sim \e^2, \quad F \sim \e \ee with $\e \to 0$ is considered so that
\be \label{limit1} M_{\m \n} = - (Bg^{-1}B)_{\m \n} , \quad
\Theta^{\m\n} = -2 \pi \a' \left(\frac{1}{F}\right)^{\m\n} \ee 
are fixed in this limit. They show that in this limit the tree level,
amplitudes which can be  computed from \eq{tree}, yields the result of
a noncommutative field theory in a metric $M_{\m\n}$ and with the
noncommutative parameter $\Th^{\m\n}$ given by \eq{limit1}.  
For example, 
\be S = \int \sqrt{\det M}~ \Big[ (\del \Phi)^2 + V_{\Th} (\Phi)
\Big].  
\ee 
Since $M$ and $\Th$ are finite in this limit, all our
multiloop formula also have a well defined limit.
On the other hand, if one prefers (see for example \cite{BCR,crs}),
one can also rescale the string coordinates by a factor $\Xh^\m = X^\n
(g -F)_\n{}^\m$ and use $\Xh$ to construct the corresponding string
amplitude.  For example, at the tree level, one gets
\be \label{treeG'}
\hat{\cG}^{\m\n}(\t,\t'):= \langle \Xh^{\m}(\t)  
\Xh^{\n}(\t')  \rangle 
= -\a' g^{\m\n} \ln (\t-\t')^2 
+ \frac{i}{2} \th^{\m\n} \e(\t-\t'),
\ee
where
\be \label{rot}
\th := (g+ F) \Theta (g-F) = 2\pi \a' F
\ee
and the metric $g$ is the closed string metric. For higher loops, one
simply has to replace everywhere (e.g. in \eq{abc}, \eq{abc0},
\eq{p1}, \eq{np1} ) $(M^{-1})^{\m\n}$, $\Th^{\m\n}$ by $g^{\m\n},
\th^{\m\n}$ in the string theoretic expressions.  Thus one can also
take the following {\it noncommutative field theory limit}
\cite{BCR,crs}
\be\label{limit2}
\a' \sim \e, \quad  \mbox{$g$ fixed}, \quad F \sim 1/\e .
\ee
so that 
\be
g_{\m \n}, \quad  \th^{\m \n} = 2\pi \a' F^{\m \n}
\ee
are fixed in the limit. In this limit, the noncommutative field theory
\be\label{26}
S = \int \sqrt{\det g}~ \Big[ (\del \Phi)^2 + V_{\th} (\Phi) \Big], 
\ee
in a background metric $g$ and with noncommutative parameter $\th$ is
resulted.  We stressed that the two 
limits are complitely
equivalent. No matter 
which limit one takes, there is always a pair
of parameters (metric and the noncommutative parameter) in the string
amplitude, and the field theory limit is always taken in such a way
that they remain finite. In the end, we obtain a field theory with a
background metric and a noncommutative parameter.

\subsection{Examples}

In what follows, we aim at producing a field theory limit with a
flat Minkowskian metric $\eta_{\m\n}$, 
which is more often considered in the literatures of noncommutative 
quantum field theory. 
This can be most easily
achieved from a string theory having a metric $g_{\m\n} = \eta_{\m\n}$
and using the rescaled $\Xh$ to construct the string amplitudes. The
field theory noncommutativity parameter $\th$ is then related to the
string moduli $F$ by equation \eq{rot}.

In order to give some concrete examples, we will now consider two
Feynman diagrams of the noncommutative scalar theory with cubic
interaction in six dimensions. We will focus first on the non--planar
contribution to the 2--point function at one loop and then on the
non--planar 2-loop vacuum bubble. Many other examples are thoroughly
described in~\cite{nn1,BCR, crs} also for scalar theories with
quartic interactions and for Yang--Mills theory.
Eq.~\eq{gluonmaster0} can be used also as a starting point for studying
scalar (i.e. tachyon) amplitudes simply by setting all the
polarizations $\epsilon_i$ to zero. If we fix $\rho_1=1$ on the first
border and the second puncture on the other border $\rho_2\in[-1,-k]$,
we have to use in the master formula the non--planar Green
function~\eq{p1}, obtaining
\be\label{adnp}
A_{2,1}^{\rm NP}(p_1,-p_1) = {\a'^{2-d/2}\over (4\pi)^{d/2}} 
{g_3^2\over 4}
\int_0^1 {d  k\over  k} \,\ex{\a' m^2 \ln{k}} \!
\left( {-\ln  k}\right)^{-d/2}
\int_{-1}^{-k}{d\r_2\over \r_2}~
\ex{\left[-2 \a' p_{1 \m} p_{1 \n} G^{\m\n}_{\rm 
NP}(1,\r_2)\right]}~,
\ee
where we have already translated the string coupling 
constant into field theory one by using the 
relation 
$g_3=2^{5/2} g_{\rm op}
(2\a')^{d-6\over 4}$, found from the matching of the 
$3$-point functions.
As usual, the logarithmic divergences in the
integrand are related to the dimensionful Schwinger 
parameters via a
factor of $\a'$. In particular, at 1-loop one always 
associates $\ln k$ to the total length of the loop by taking,
$\ln  k = -{T/\a'}$
where $T$ has to be kept finite as $\a'\to 0$. 
In this limit $k$ goes to zero exponentially
which means shrinking the annulus to a one-loop Feynman 
graph. 
After having replaced string
quantities with field theory ones, the $\a'$ dependence of
(\ref{adnp}) simplifies and the whole amplitude is just 
proportional to a single power of $\a'$. This means that, in order to 
have a finite answer for our field theory limit, it is necessary to 
introduce one more Schwinger parameter. Since we want to reproduce the 
irreducible diagram, we perform the $\a'\to 0$ limit by keeping fixed 
also $t_2$ $\ln \r_2=-{t_2/\a'}$ and get 
\be\label{adnpft}
A_{2,1}^{\rm NP}(p_1,-p_1) = {g_3^2\over 4}{1\over 
(4\pi)^{d/2}} 
\int\limits_0^\infty {dT\over T^{d/2}} {\rm e}^{-m^2 T}
\int\limits_0^T dt_2 
\exp{\left[ -p_1^2 t_2 \left( 1-{t_2\over
T}\right)+  p_1^\m p_1^\n {\th^2_{\m\n}\over 4T} \right]} 
\ .
\ee
Thus we recover the standard Schwinger
proper time integral, but with the additional factor 
typical of the noncommutative diagrams.
%
As noted in~\cite{uvir}, for $\pt^2\ne 0$ this serves as an
effective UV cutoff and is at the origin of the UV/IR mixing. 

In the above example the non-planarity of the diagram is related to
the insertion of the external legs. From the string point of view this
follows from the fact that the two insertions are done on the two
different borders of the annulus. In field theory this means that the 
two 3-point vertices in the graph have different cyclical 
orientation (and so the Filk phase has opposite sign). At one loop,
this kind of non--planarity is the only possible one. As we saw in the 
previous string calculation, at multiloop level a new feature 
appears, namely the internal legs can intersect and bring a
non--trivial modification of the string measure~\eq{stringmaster}. A
first simple check of this string result is to show that, at low
energies, it can reproduce the result of noncommutative Feynman
diagrams.

The simplest field theory displaying this feature is the irreducible 
vacuum bubble that is obtained by sewing together two $3$--point 
vertices with different relative orientations. As in the above 
$1$--loop diagram, the two Filk factors combine and give an additional
phase factor ${\rm e}^{\ii r\theta q}$ with the respect to the usual 
computation. However, now we have to integrate over both momenta 
contracted with $\theta$. Carrying out the $r$ integral, we obtain
\be
A^{\rm NP}_{0,2} = \int dq \int \frac{dt_i }{l^{d/2}} {\rm e}^{-m^2
  \sum_i t_i}\;
\exp \left( -\frac{1}{l}[q^\m \Dth_{\m\n} q^\n]\right),
\ee
where $t_i$ are the Schwinger parameters related to the three 
propagators, $l := t_2+t_3$. Finally $\Dth$ is the following diagonal 
matrix 
\be
(\Dth)_{\m\n}:=\d_{\m\n} (t_1 t_2 + t_1 t_3 +t_2 t_3) - 
\frac{\th^2_{\m\n} }{4}~,
\ee
which matches exactly  the leading contribution in the 
multiplier of the expression for $A^{\mu\nu}_{IJ}$ in \eq{abc}, once 
the string parameters have been translated in Schwinger proper times 
\cite{crs}.

\section{Some Remarks on UV/IR}

In this paper, we studied open string amplitudes in the presence of a
constant $B$-field. We derived from first principles a closed
expression~\eq{stringmaster} for the $N$-Reggeon vertex encoding the
presence of the non-trivial 
background. From this Reggeon vertex one can 
generate all string loop diagrams by the usual techniques.
Then we considered the field theory limits (\eq{limit1} and
\eq{limit2}) which, at disk level, give rise to the 
star-product and to noncommutative Lagrangians. In the same limits, the
multiloop open string amplitudes derived from the new Reggeon vertex
reduces to the loop diagrams of a noncommutative field
theory. Moreover, as in the commutative case, the matching of the
string and the field theory calculations is evident at the level of
integrands and is diagram by diagram. These observations are further
evidence that non commutative field theories can be embedded in string
theory in a consistent way. 

However within this point of view, there is
still one important point which deserves a better understanding, that
is the UV/IR mixing~\cite{uvir}. In fact, if string theory provides a
microscopical definition for noncommutative field theories, then in
principle one should be able to apply Wilson approach and obtain a low
energy action describing the physics in the IR. However, in the
noncommutative case, one typically finds that (for $\tilde{p}^2$)
there are massless poles which do not correspond to any degree of
freedom present in the effective action.

There have been various different proposals to explain this
puzzle from a string point of view. A possibility is that other string
states, besides those of open strings, are not compltely decoupled
either because in the non planar diagrams open strings can never be
really point-like~\cite{hong} or because closed strings play a
relevant role \cite{uvir}, see also \cite{shenker}. 
However, even if this is true, it has not been possible
so far to identify clearly the nature of the states entering in the
effective action.
Another possibility is that the UV/IR is just an artifact of the
perturbative expansion. For instance, the pure scalar field theory we
considered can be embedded easily in 
the bosonic open string theory, which is tachyonic 
(here we expanded around the usual flat vacuum). 
However, it is by now clear that open string tachyon can
condense giving rise to stable vacua and it is possible that a field
theory limit around these new vacua is not pathological. 
On the other hand UV/IR mixing can also be 
presented in supersymmetric gauge theories  which can be embedded 
in perfectly consistent string models. Perturbative analysis of 
supersymmetric gauge theory with focus 
on aspects of UV/IR can be found in \cite{susygauge}. 
From the knowledge accumulated on the commutative gauge theory, 
it is perhaps not surprising that nonperturbative contribution may 
modifies substantially  the physics in the IR and may provide a 
cure to the UV/IR mixing. 
Recently the case of  noncommutative pure $N=2$ Yang-Mills theory has
been analyzed beyond the perturbative level \cite{nonpert0}, 
and particularly in \cite{nonpert1}.
It seems that the contribution of instantons
makes the IR behvior less pathological.
 
\vskip0.5cm
\noindent
{\large \bf Acknowledgments}

\smallskip
\noindent

The results presented here are based on~\cite{BCR,crs}. We would like 
to thank A. Bilal and S. Sciuto for their collaboration.
This work has been supported by the Swiss National Science
Foundation, by the European Union under RTN contract 
HPRN-CT-2000-00131.



\begin{thebibliography}{10}

\bibitem{CH1}
C.-S. Chu and P.-M. Ho,
\newblock Nucl. Phys. {\bf B550}, 151 (1999);
\newblock {\tt hep-th/0001144}.

\bibitem{crs}
C.-S. Chu, R.~Russo, and S.~Sciuto,
\newblock Nucl. Phys. {\bf B585}, 193--218 (2000).

\bibitem{dolan}
L.~Dolan and C.~R. Nappi,
\newblock {\tt hep-th/0009225}.

\bibitem{dss}
S.~Sciuto,
\newblock Nuovo Cimento Lett. {\bf 2}, 411 (1969). \\
\newblock P. Di Vecchia,  R. Nakayama, J.L. Petersen, S. Sciuto,
\newblock Nucl. Phys. {\bf 282} (1987) 103.


\bibitem{sc2}
P.~Di Vecchia, M.~Frau, A.~Lerda, and S.~Sciuto,
\newblock Phys. Lett. {\bf B199}, 49 (1987).


\bibitem{filk}
T.~Filk,
\newblock Phys. Lett. {\bf B376}, 53 (1996).

\bm{russo1}  P. Di Vecchia, A. Lerda, L. Magnea, R. Marotta, R. Russo,
Nucl. Phys. {\bf B469} (1996) 235.

\bibitem{phi}
P.~Di Vecchia, F.~Pezzella, M.~Frau, K.~Hornfeck, A.~Lerda, and S.~Sciuto,
\newblock Nucl. Phys. {\bf B322}, 317 (1989).

\bibitem{frau1997}
M.~Frau, I.~Pesando, S.~Sciuto, A.~Lerda and R.~Russo,
Phys.\ Lett.\ B {\bf 400}, 52 (1997)

\bm{callan} A. Abouelsaood, C.G. Callan, C.R. Nappi, S.A. Yost,
Nucl. Phys. {\bf B280} (1987) 599.

\bibitem{Andreev}
O.~Andreev,
Phys.\ Lett.\ B {\bf 481}, 125 (2000)

\bibitem{nn1}
O.~Andreev and H.~Dorn,
\newblock Nucl.\ Phys.\ {\bf B583} (2000) 145.

\bibitem{nn2}
Y. Kiem and S. Lee,
\newblock Nucl.\ Phys.\ {\bf B586} (2000) 303.

\bibitem{BCR}
A.~Bilal, C.~Chu and R.~Russo,
Nucl.\ Phys.\ {\bf B582} (2000) 65.

\bibitem{shenker}
J. Gomis, M. Kleban, T. Mehen, M. Rangamani, and S. Shenker,
\newblock JHEP {\bf 08}, 011 (2000).

\bibitem{hong}
H. Liu and J. Michelson,
\newblock Phys. Rev. {\bf D62}, 066003 (2000).

\bibitem{sw1}
N.~Seiberg and E.~Witten,
\newblock JHEP {\bf 09}, 032 (1999).

\bm{kiem}  Y. Kiem, S. Lee, J. Park, 
Nucl. Phys. {\bf B594} (2001) 169.

\bibitem{scherk71}
J.~Scherk,
\newblock Nucl. Phys. {\bf B31}, 222 (1971).

\bibitem{fmr}
Alberto Frizzo, Lorenzo Magnea, and Rodolfo Russo,
\newblock Nucl.\ Phys.\ {\bf B579} (2000) 379.

\bibitem{mp}
Raffaele Marotta and Franco Pezzella,
\newblock Phys.\ Rev.\ D {\bf 61}, 106006 (2000).

\bibitem{fmr2}
Alberto Frizzo, Lorenzo Magnea, and Rodolfo Russo,
\newblock {\tt hep-ph/0012129}.

\bibitem{bddpr}
Z.~Bern, L.~Dixon, D.~C. Dunbar, M.~Perelstein, and J.~S. Rozowsky,
\newblock Nucl. Phys. {\bf B530}, 401 (1998).


\bibitem{uvir}
S.~Minwalla, M.V. Raamsdonk, and N.~Seiberg,
\newblock {\tt hep-th/0002186},
\newblock {\tt hep-th/9912072}.

\bm{susygauge}  
H. Liu, J. Michelson, {\tt hep-th/0008205, hep-th/0011125}; \\
M.T. Grisaru, S. Penati, {\tt  hep-th/0010177}; \\
D. Zanon,  {\tt hep-th/0010275, hep-th/0011140, hep-th/0012009 };\\
V.V. Khoze, G. Travaglini, {\tt hep-th/0011218}.

\bm{nonpert0} 
M.M. Sheikh-Jabbari. {\tt hep-th/0001089};\\
K. Yoshida, {\tt hep-th/0009043};\\
D. Bellisai, J.M. Isidro, M. Matone, {\tt hep-th/0009174}.


\bm{nonpert1} 
A. Armoni, R. Minasian, S. Theisen,  {\tt hep-th/0102007};\\
T.J. Hollowood, V.V. Khoze, G. Travaglini,  {\tt hep-th/0102045}. 

\end{thebibliography}

\end{document}